# Tumbling motion of 1I/'Oumuamua reveals body's violent past*

M. Drahus[1*], P. Guzik[1], W. Waniak[1], B. Handzlik[1], S. Kurowski[1], S. Xu[2]

[1] Jagiellonian University, Kraków, Poland (drahus@oa.uj.edu.pl)

[2] Gemini Observatory, Hilo, HI, USA

**Models of the Solar System evolution show that almost all the primitive material leftover from the formation of the planets was ejected to the interstellar space as a result of dynamical instabilities[1]. Accordingly, minor bodies should also be ejected from other planetary systems and should be abundant in the interstellar space[2]. The number density of such objects, and prospects for their detection as they penetrate through the Solar System, were speculated about for decades[3,4], recently rising high hopes with the Pan-STARRS[5,6] and LSST[7] surveys. These expectations materialized on 18 October 2017 with the Pan-STARRS's discovery of 1I/'Oumuamua[8]. Here we report homogeneous photometric observations of this body from Gemini North, which densely cover a total of 8 hr over two nights. A combined ultra-deep image of 1I/'Oumuamua shows no signs of cometary activity, confirming the results from earlier, less sensitive searches[9–12]. Our data also show an enormous range of brightness variations > 2.5 mag[9], larger than ever observed in the population of Solar System objects, suggesting a very elongated shape of the body. But most significantly, the light curve does not repeat exactly from one rotation cycle to another and its double-peaked periodicity of 7.5483±0.0073 hr from our data is inconsistent with earlier determinations[9,10,13–15]. These are clear signs of a tumbling motion, a remarkable characteristic of 1I/'Oumuamua's rotation, consistent with a catastrophic collision in the distant past. This first example of an impacted minor body of exosolar origin indicates that collisional evolution of minor body populations in other planetary systems is not uncommon.**

1I/'Oumuamua[8] is a newly discovered minor body on a strongly hyperbolic trajectory, which starkly contrasts with even the most extreme cometary orbits bound to the Sun (Figure 1) and can only be explained by having origin far in the interstellar space[8,16,17]. Unsurprisingly, the Galactic velocity vector of this object is consistent with the median velocity vector of nearby stars[17]. 1I/'Oumuamua entered the Solar system from the direction of the Solar apex with a large hyperbolic excess speed of 26 km s$^{-1}$, then reached perihelion at 0.26 au from the Sun on 9 September 2017, approached the Earth to 0.16 au on 16 October 2017, and is currently on its way back to the interstellar space. Previous studies have not detected any signs of cometary activity[9–12], found optical colors consistent with Solar System's C-type and D-type asteroids[10,13,14] and reported an average V-band absolute magnitude of 22.8 to 23.0[10,12]. Such an intrinsically faint body could be discovered only thanks to the close Earth flyby in October 2017, which created a tremendous opportunity for the first detailed characterization of a minor object of exosolar origin.

We were awarded Director's Discretionary Time on the 8.1 m Gemini North telescope (program GN-2017B-DD-7) to observe this unique object. On 27 and 28 October 2017 we obtained a total of 442 r'-band[18] images using a 30 sec integration time. The sky was photometric and median seeing close to 0.6 arcsec in FWHM throughout the observations. We took data with the Gemini Multi-Object Spectrograph (GMOS), which consists of three adjacent Hamamatsu CCDs, providing imaging over a 5.5×5.5 arcmin field of view[19]. The instrument was configured to provide 2×2 binning and 0.1614 arcsec effective pixels.

The images were corrected for overscan, bias, and flatfield in the standard manner. Then, using our established technique[20], we subtracted background stars and galaxies interfering with 1I/'Oumuamua as they moved across the field of view. As a result, we obtained uncontaminated images of the target on a very clean and uniform background, suitable for accurate time-resolved photometry (see Methods). The frames were then visually inspected for artifacts, such as cosmic ray hits and residual signal from imperfect subtraction of very bright background objects at a location of 1I/'Oumuamua. This procedure resulted in the rejection of 11 images, a small number compared to the remaining 431 images available for analysis. The restricted dataset was photometrically measured using the Aperture Photometry Tool[21] configured in median sky subtraction mode. Brightness of 1I/'Oumuamua was determined from a circular aperture of a constant 2.26 arcsec (14 pixels) diameter, which ensured negligible influence of seeing variations and acceptable level of background noise contribution. We also made similar measurements of brighter background stars, employing this time a larger aperture of 4.8 arcsec (30 pixels) to account for the elongation of stellar profiles caused by rapid non-sidereal tracking of the target. Differential photometry of 1I/'Oumuamua was then reduced to the observing geometry on the first night and absolute calibrated through the background stars available in the SDSS Photometric catalog[22].

In Figure 2 we present a mean combined image of 1I/'Oumuamua, having an effective integration time of 214.5 min. No signs of cometary activity can be seen, consistent with earlier reports[9–12], however, the sensitivity of our image is a factor of 2.6 to 9.5 higher than those achieved by the previous searches. Our light curve of 1I/'Oumuamua densely covers a total of 8 hr over two nights. The median flux corresponds to an observed r'-band magnitude of 22.25 and an absolute magnitude of 22.02, the latter calculated in the standard way assuming the asteroidal photometric phase function[23] with a parameter $G = 0.15$ characteristic of solar-system C-types[24]. Our absolute magnitude is broadly consistent with the absolute magnitudes reported before[10,14]. More interestingly, the range of brightness variations is enormous, at least a factor of ten, or > 2.5 mag, which exceeds even the largest rotational variations observed in the entire population of 16,419 solar-system asteroids with measured light curves (Figure 1). The light curve of 1I/'Oumuamua was scrutinized for periodicity using the classical Phase Dispersion Minimization algorithm[25] with implemented inverse-variance weighting of the data points[26]. Periodograms (see example in Figure 3) systematically show the best data phasing for a frequency of 0.13248±0.00013 hr$^{-1}$, or 7.5483±0.0073 hr period, implying a double-peaked phased light curve (Figure 4), most easily explained by shape-dominated brightness variations. Though not as good, a corresponding single-peaked phasing, consistent with albedo-dominated brightness variations, is also clearly indicated by the periodograms at approximately twice the double-peaked frequency, 0.26596±0.00026 hr$^{-1}$, or half the double-peaked period, 3.7599±0.0037 hr. Other periodicities offer implausible data phasings and can be safely excluded. We proceed assuming shape-dominated brightness variations and, consequently, adopt the double-peaked periodicity as 1I/'Oumuamua's synodic rotation period. From Hapke modeling of the light curve[27] we find the long-to-short axis ratio of > 4.63 and the effective radius of about 80 m (see Methods). The very large elongation together with the measured moderate rotation rate require a density of > 1034 kg m$^{-3}$ to prevent the body from falling apart. This limit is calculated under the assumption that the tensile strength is negligible, which may or may not be true for 1I/'Oumuamua. Nonetheless, our estimate shows that the body may be strengthless and still have a density within the ranges of typical Solar System asteroids[28], in contrast with the previous revelations[9,10,14].

While the light curve of 1I/'Oumuamua is clearly periodic, it does not repeat exactly from one rotation cycle to another. As we cannot explain this behavior by instrumental effects, we conclude that it is a real feature of the light curve, intrinsic to the object. Furthermore, the light curve does not appear to have a single, unique periodicity because the rotation periods reported by other studies[9,10,13–15] differ from one another and are inconsistent with our data (Figure 5). We recognize these peculiarities as the characteristic signatures of non-principal-axis (or excited) rotation, also often referred to as tumbling[29,30], which has significant consequences for understanding the distant history of this object.

Although the vast majority of Solar System minor bodies do not show any measurable deviations from simple rotation, a small fraction of asteroids[29,31] and a few comets[32,33] have been identified as tumblers. In particular, comet 1P/Halley was the first minor body found in non-principal-axis rotation state[32] and, subsequently, asteroid (4179) Toutatis was identified as the first tumbler among asteroids[34]. Rotational excitation occurs mainly through collisions and, restricted to comets, sublimation torques. A tumbler then dissipates rotational energy due to stresses and strains resulting from complex rotation, and returns to simple, minimum-energy rotation state on a certain timescale. The damping timescale depends on the object's size, shape, density, stiffness, and rotation rate, and ranges from hundreds years to hundreds of billions years for the known Solar System asteroids[35].

Detailed quantification of 1I/'Oumuamua's complex rotation state is beyond the scope of this paper. Here we explore the fundamental fact that the body was once excited and has not fully relaxed yet. Evidently, the excitation is not easily explainable by sublimation torques. First, 1I/'Oumuamua does not show any signs of active outgassing despite a superb sensitivity of our combined image (Figure 2) and exceptionally favorable orbital circumstances, and second, it is unlikely that the body was active in the past because the sublimation levels required to excite its rotation would also generate enormous changes in the rotation rate, quickly leading to rotational instability and disruption. Instead, the complex rotation state of 1I/'Oumuamua most probably originates from an impact. Collisional excitation of this body in our Solar System is hardly possible, though, because of the rarity of collisions even in the (not so) dense main asteroid belt[36] and because 1I/'Oumuamua missed the belt at a safe distance and spent very little time close to the ecliptic plane due to the highly inclined orbit and the large orbital speed. A sensitive non-detection of a body's debris trail (Figure 2), starkly contrasting with the prominent trail of the similarly-sized impacted asteroid P/2010 A2[37], is also not in favor of a very recent collision. Rather, we believe that 1I/'Oumuamua was excited in another planetary system – presumably its home system – in the distant past. A damping time scales of typical rubble pile asteroids having the same effective size and rotation rate as 1I/'Oumuamua, and body's minimum allowable axis ratio, is of about 1 Gyr[38]. This appears to be long enough to preserve the signs of collisional excitation over the timescale of a typical interstellar exile[11], supporting our conclusion. Whether 1I/'Oumuamua was excited via a giant collision, giving birth to this object, or impacted by another small body, is unclear, but both scenarios lead to the same conclusion that collisional processing of minor body populations in exoplanetary systems is not uncommon. Supposedly, 1I/'Oumuamua was ejected from its home system with a large number of similar bodies during a period of dynamical instability[1] when frequent collisions are expected.

**Acknowledgements**

M.D., P.G. and B.H. are grateful for support from the National Science Centre of Poland through a SONATA BIS grant no. 2016/22/E/ST9/00109. We thank the staff of the Gemini Observatory for assistance and are indebted to the Director of Gemini Observatory for allocating Gemini North time for this program.

## Competing Interests

The authors declare that they have no competing financial interests.

## Methods

**Image cleaning.**

In order to obtain accurate time-resolved photometry of 1I/'Oumuamua, it was necessary to carefully subtract a dense background of stars and galaxies along its trajectory. To remove the background from a given image, we mean combined up to six background-equalized and flux-normalized images taken around the same time and subtracted the result of this operation from the reference image. We did not use the nearest images from a consecutive series to avoid contaminating the photometric background annulus with the object's signal. An example of the application of this technique to our 1I/'Oumuamua data is presented in Figure S1.

**Light curve modeling.**

We synthesized the light curve of 1I/'Omuamua with the Hapke approach[27], using MIMSA (modified isotropic multiple scattering approximation) with SHOE (shadow-hiding opposition effect) and macroscopic roughness influencing reflectance properties. Colorimetric and spectroscopic observations have revealed that the reflectance spectrum of this object is close to those of the Solar System C- or D-type asteroids, Trojans, or comets[10,11,13,39]. We adopted in our model a set of reasonable literature values for the reflectance parameters of C-type asteroids[40], such as a single scattering albedo of 0.037, an opposition surge amplitude and width of 0.20 and 0.025, respectively, the asymmetry factor in the Henyey-Greenstein particle phase function equal to -0.47, as well as an average topographic slope angle of macroscopic roughness of 20°. For simplicity, we assumed that 1I/'Oumuamua has a shape of a prolate spheroid. To test how the solution depends on the spin axis orientation we probed a number of directions with respect to the line of sight (aspect angle θ) and the Sun-object-observer plane (polar angle φ measured counterclockwise from the object-Sun direction as seen from observer position). This orientation has been taken for the geometry of sight on 28 October 2017, 00:00 UT and fixed in the external, inertial frame for our whole observing run.

We have checked how the peak-to-peak amplitude of the light curve depends on the axial ratio for different orientations of the spin pole. A dependency of the flux ratio versus the axial ratio is presented in Figure S2. Even an enormous amplitude of the light curve does not necessarily mean a huge axial ratio. We found that to generate a 2.5-mag amplitude, an axial ratio can be as low as 4.63, inconsistent with the recently published values[9,10,13]. Large axial ratios were reported with the assumption that the brightness variation comes purely from cross-section changes. Our lower limit is a consequence of the assumed reflectance properties of the asteroid surface and partial shadowing due to over 20° phase angle. When a phase angle is far from 0° and the body is elongated, shadowing effects as well as asymmetry of the phase function have a strong influence on an observed light curve. Moreover, for aspect angles smaller than 60° it is virtually impossible to obtain a peak-to-peak amplitude consistent with 1I/'Oumuamua in spite of blowing up the axial ratio to abnormal values. Thus we conclude that aspect angle must be larger than 60°.

# Figures

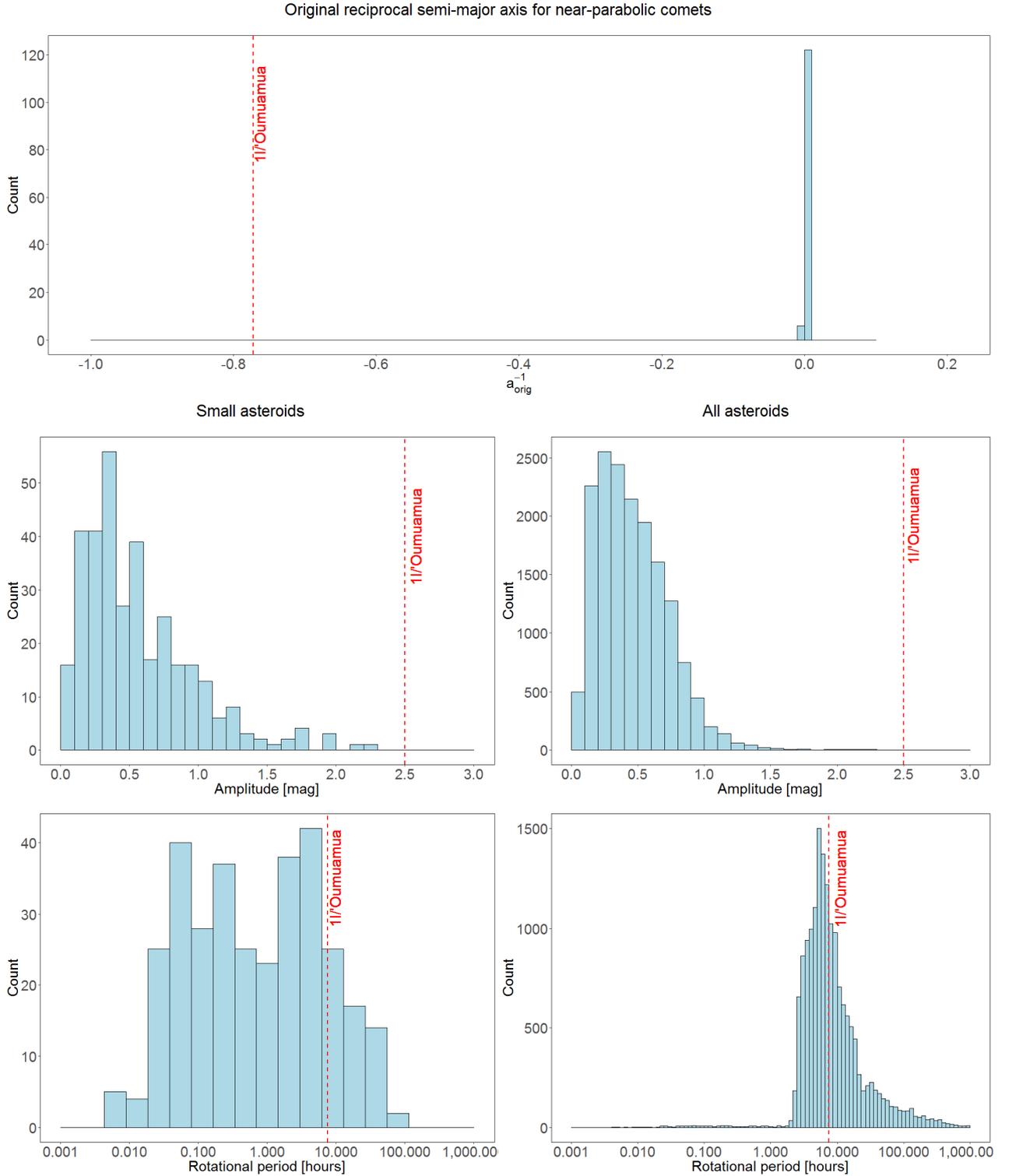

**Figure 1.** 1I/'Oumuamua in the context of known minor Solar System bodies. *Top:* Distribution of the original reciprocal semi-major axes of the near-parabolic comets[41]. *Middle:* Distribution of the light curve amplitudes of small (*left*) and all (*right*) asteroids[31]. *Bottom:* Distribution of the rotation periods of small (*left*) and all (*right*) asteroids[31]. Small asteroids are defined as objects with diameters < 200 m.

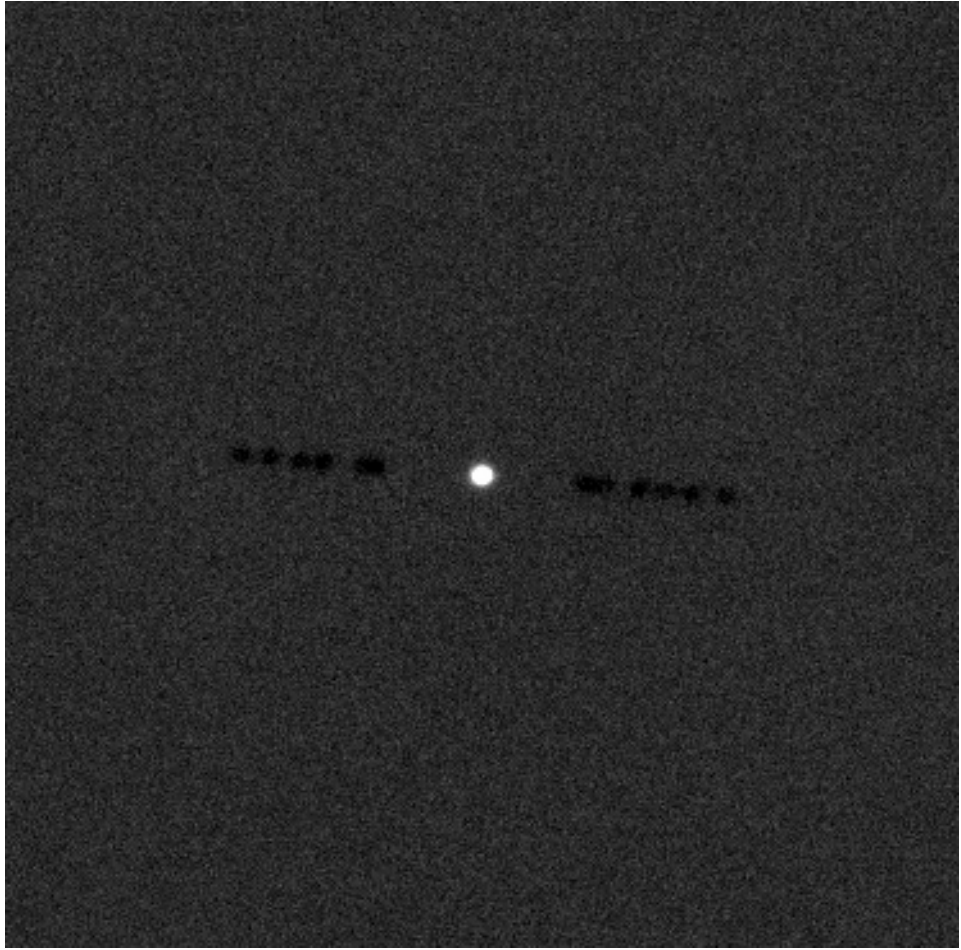

**Figure 2.** Deep stack of our r'-band time series, having an effective integration time of 214.5 min. The negative images of the target to the left and right of the actual positive image were produced by our background subtraction algorithm and do not affect the photometry. The presented region is 5.5×5.5 arcmin, North is to the top and East to the left. Despite having a very high surface brightness sensitivity of 28.1 mag arcsec$^{-2}$ measured in a 1 arcsec$^2$ region, the image does not show any signs of cometary activity.

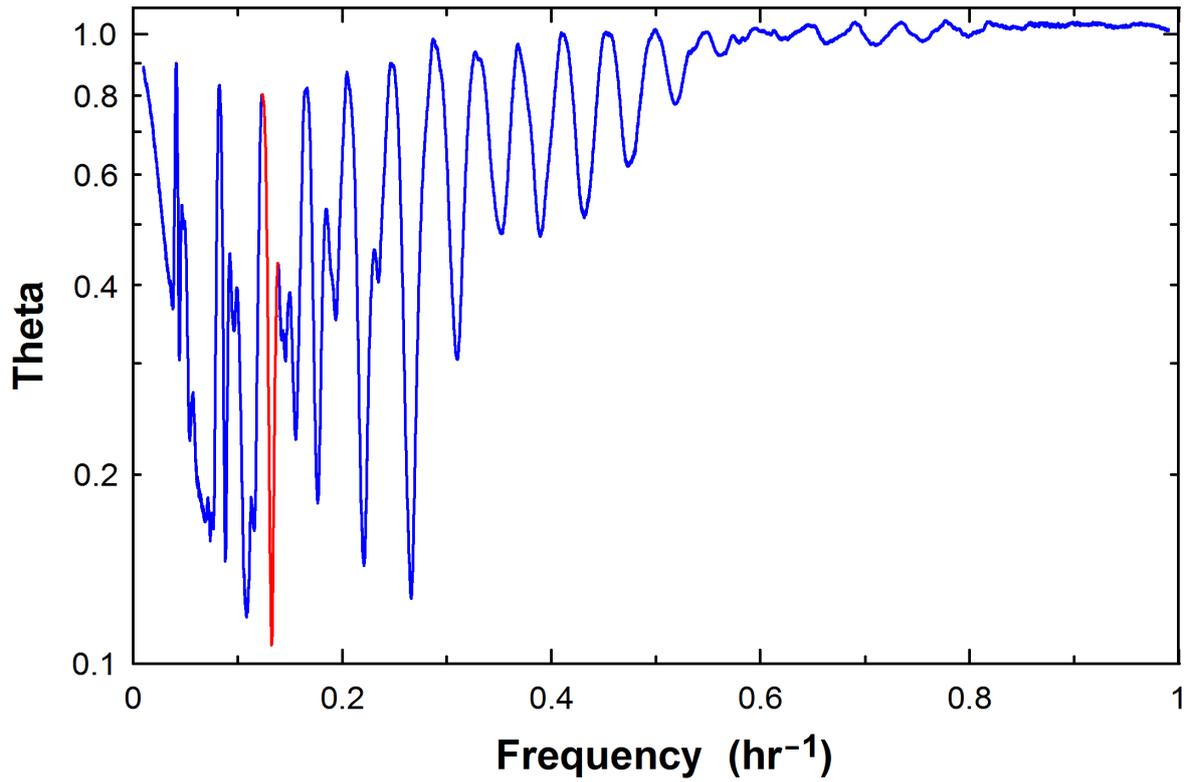

**Figure 3.** Phase Dispersion Minimization periodogram calculated with 25 bins and 5 covers for our time series[25,26]. Red indicates the best periodicity solution for a frequency of 0.13248±0.00013 hr$^{-1}$, or 7.5483±0.0073 hr period, consistent with a double-peaked phased light curve.

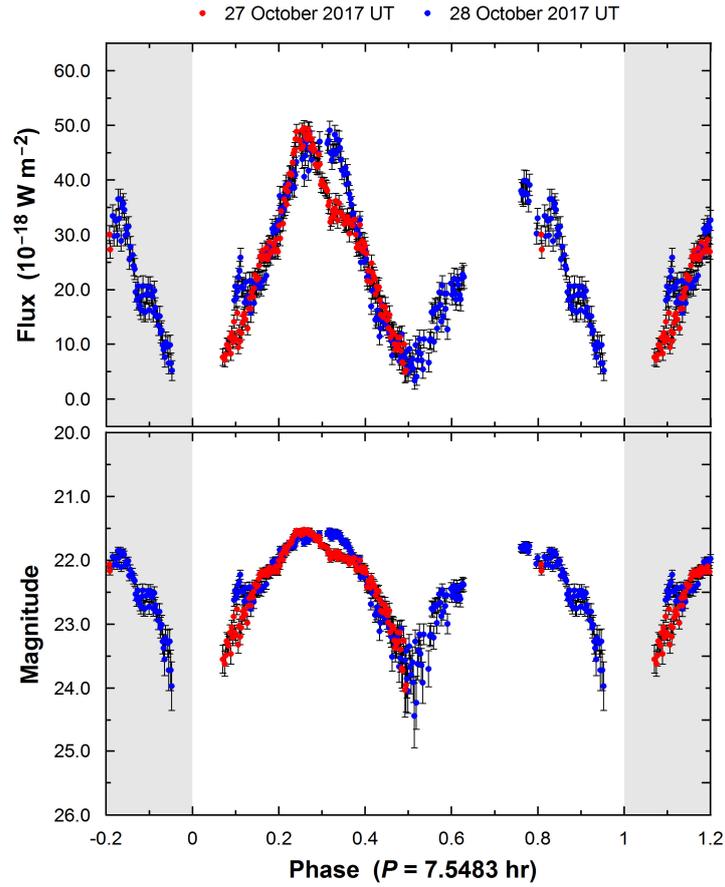

**Figure 4.** Our data phased for the best periodicity solution, $P_{rot}$ = 7.5483 hr. It is evident that the light curve does not repeat exactly from one night (27 October 2017 UT) to another (28 October 2017 UT).

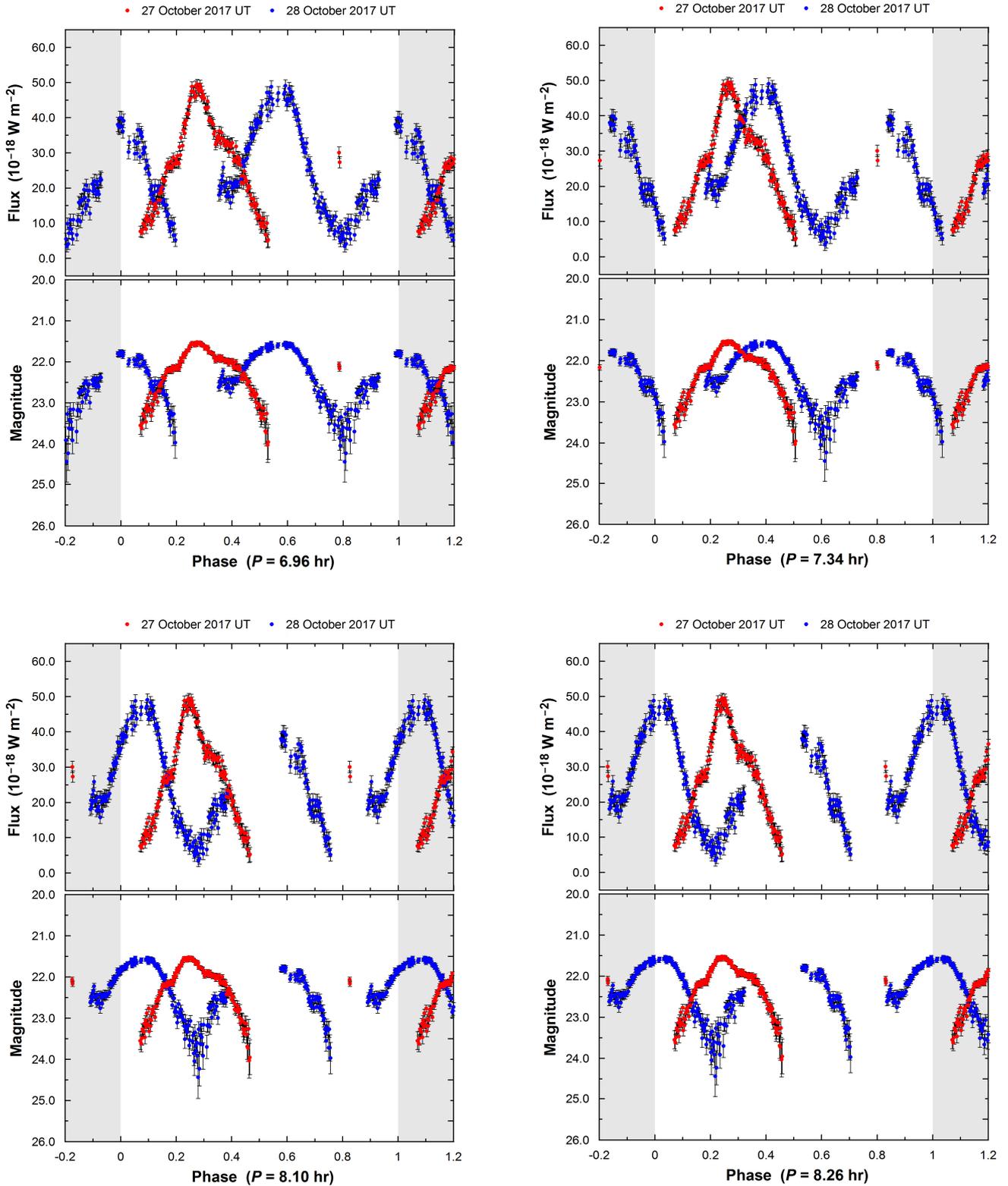

**Figure 5.** The same as before but phased for 6.96 hr found by Feng et al.[15] (*top-left*), 7.34 hr found by K. Meech et al.[9] (*top-right*), 8.10 hr found by B. Bolin et al.[13] and M. Bannister et al.[14] (*bottom-left*), and D. Jewitt et al.[10] (*bottom-right*). None of these four periodicity solutions phase our data, showing that 1I/'Oumuamua does not have a unique rotation period, as expected from tumbling motion.

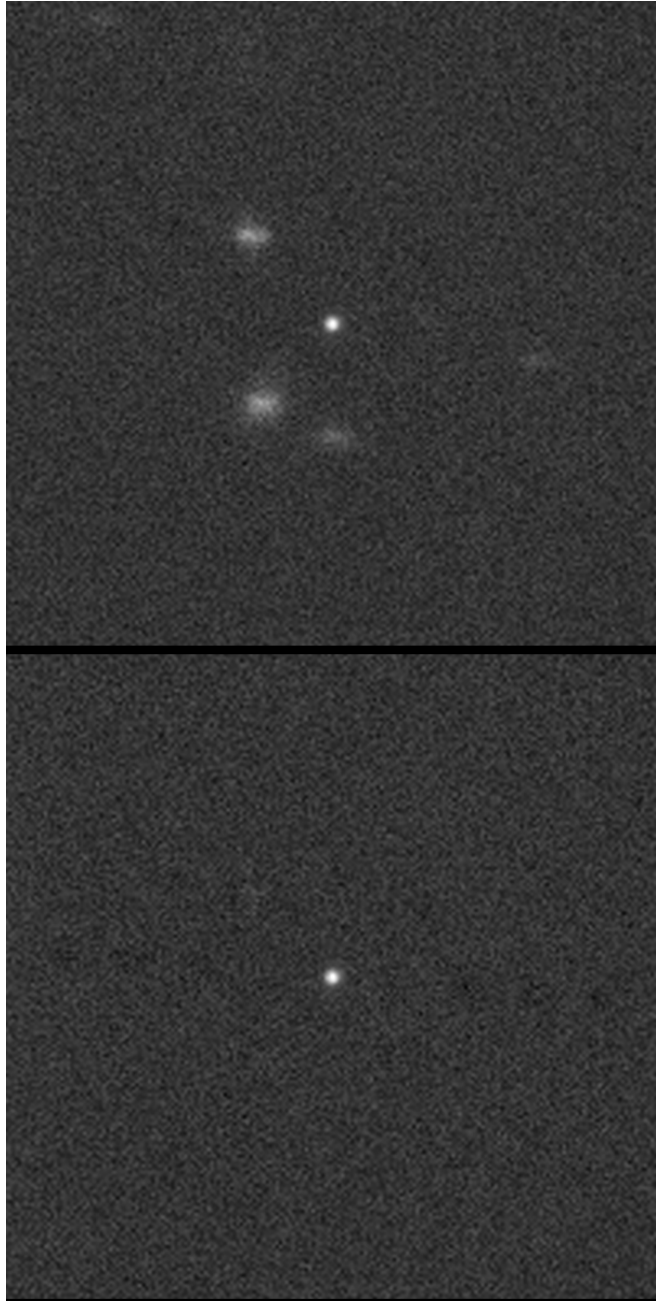

**Figure S1.** Demonstration of our background cleaning algorithm. *Top:* example of an original image, corrected only for overscan, bias, and flatfield. *Bottom:* clean version of the image above, with stars and galaxies accurately removed.

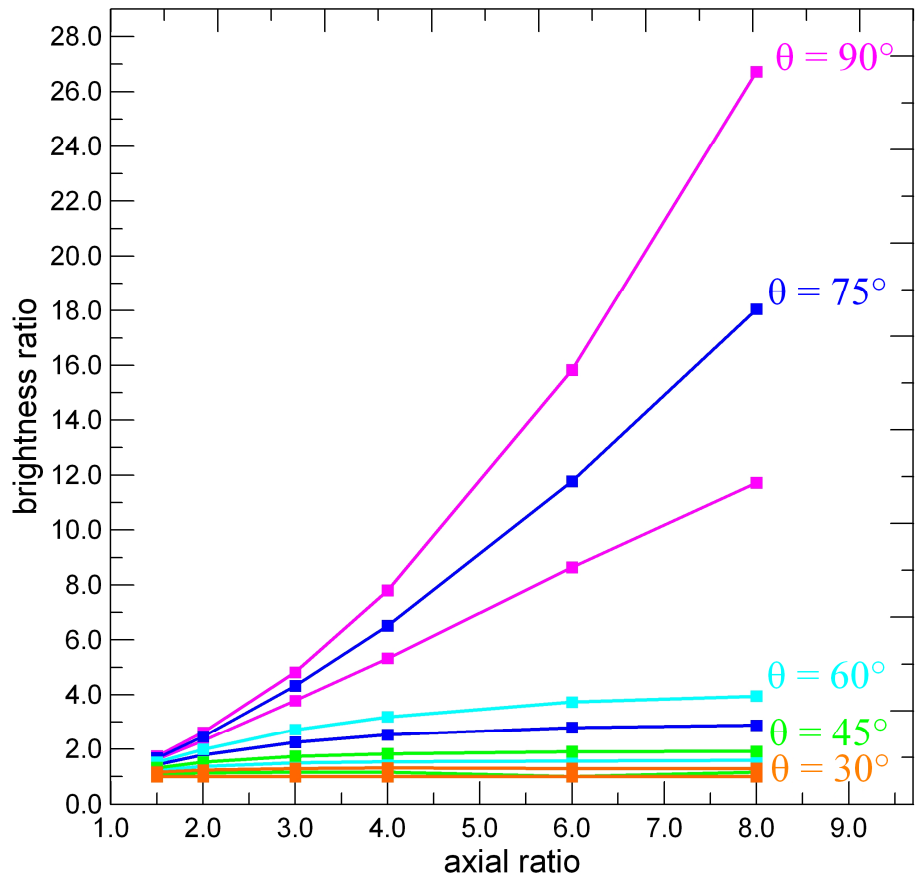

**Figure S2.** Flux ratio versus axial ratio for aspect angle θ from 90° to 30°. For each angle two different trajectories are presented, one for minimum and one for maximum flux ratios over a series of polar angles φ for a given aspect angle θ.

## Supplementary References